\documentclass[conference]{IEEEtran}
\IEEEoverridecommandlockouts

\usepackage{amsmath,amsfonts}
\usepackage[utf8]{inputenc}
\usepackage{algorithm}
\usepackage{array}
\usepackage{algpseudocode}
\usepackage{todonotes}
\usepackage{graphicx}
\usepackage{textcomp}
\usepackage{svg}
\usepackage{xcolor}
\usepackage{longtable}
\usepackage{listings}
\usepackage{multirow}
\usepackage{lineno}
\def\BibTeX{{\rm B\kern-.05em{\sc i\kern-.025em b}\kern-.08em
    T\kern-.1667em\lower.7ex\hbox{E}\kern-.125emX}}
    
\usepackage{flushend}
\usepackage{tcolorbox}
\usepackage{color}
\usepackage{xspace}
\usepackage{amssymb}
\usepackage{tabularx}
\usepackage{booktabs}
\usepackage{longtable}
\usepackage[font=small,skip=0pt]{caption}
\usepackage{subcaption}
\usepackage{url}
\usepackage{tikz}
\usepackage{enumitem}
\usepackage{adjustbox}

\usepackage{tcolorbox}

\newcommand{\findingbox}[2]{\begin{tcolorbox}[boxsep=3pt,left=4pt,right=4pt,top=2pt,bottom=2pt,colback=gray!20,colframe=gray!80!black,title={#1}]
{#2}
\end{tcolorbox}}

\usepackage{hyperref}

\usepackage{makecell}
\usepackage{pifont}%
    
\newboolean{showcomments}
\setboolean{showcomments}{true}
\ifthenelse{\boolean{showcomments}}
 { \newcommand{\mynote}[2]{
      \fbox{\bfseries\sffamily\scriptsize#1}
        {\small$\blacktriangleright$\textsf{\emph{#2}}$\blacktriangleleft$}}}
        { \newcommand{\mynote}[2]{}}

\newcommand{\toolname}{{\it ArkAnalyzer}\xspace}
\newcommand{\OH}{{OpenHarmony}\xspace}
\newcommand{\arkts}{{ArkTS}\xspace}

\definecolor{codegreen}{rgb}{0,0.6,0}
\definecolor{codegray}{rgb}{0.5,0.5,0.5}
\definecolor{codepurple}{rgb}{0.58,0,0.82}
\definecolor{backcolour}{rgb}{0.95,0.95,0.92}

\definecolor{lightgray}{rgb}{.9,.9,.9}
\definecolor{darkgray}{rgb}{.4,.4,.4}
\definecolor{purple}{rgb}{0.65, 0.12, 0.82}

\lstdefinelanguage{JavaScript}{
  keywords={typeof, new, true, false, catch, function, return, null, catch, switch, var, if, in, while, do, else, case, break},
  keywordstyle=\color{blue}\bfseries,
  ndkeywords={class, export, boolean, throw, implements, import, this},
  ndkeywordstyle=\color{darkgray}\bfseries,
  identifierstyle=\color{black},
  sensitive=false,
  comment=[l]{//},
  morecomment=[s]{/*}{*/},
  commentstyle=\color{purple}\ttfamily,
  stringstyle=\color{red}\ttfamily,
  morestring=[b]',
  morestring=[b]"
}

\usepackage{xspace}
\makeatletter
\DeclareRobustCommand\bmvaOneDot{\futurelet\@let@token\bmv@onedotaux}
\def\bmv@onedotaux{\ifx\@let@token.,\else.\null\fi\xspace}
\DeclareRobustCommand\bmvaTwoDot{\futurelet\@let@token\bmv@twodotaux}
\def\bmv@twodotaux{\ifx\@let@token.,\else.,\null\fi\xspace}
\makeatother
\def\eg{\emph{e.g}\bmvaTwoDot} 
\def\ie{\emph{i.e}\bmvaTwoDot}

\begin{document}

\title{\toolname: The Static Analysis Framework for \OH}

\author{\IEEEauthorblockN{Haonan Chen, Daihang Chen, \\Yizhuo Yang}
\IEEEauthorblockA{School of Software \\
Beihang University\\
China}
\and
\IEEEauthorblockN{Lingyun Xu, Liang Gao}
\IEEEauthorblockA{CBG Software Engineering \\Department\\Huawei\\
China}
\and
\IEEEauthorblockN{Mingyi Zhou, Chunming Hu,\\ Li Li$^{\ast}$ \thanks{*Corresponding author}}
\IEEEauthorblockA{School of Software \\
Beihang University\\
China}
}

\maketitle

\begin{abstract}
\arkts is a new programming language dedicated to developing Apps for the emerging \OH mobile operating system.
Like other programming languages (e.g., Typescripts) constantly suffering from performance-related code smells or vulnerabilities, the \arkts programming language will likely encounter the same problems.
The solution given by our research community is to invent static analyzers, which are often implemented on top of a common static analysis framework, to detect and subsequently repair those issues automatically.
Unfortunately, such an essential framework is not available for the \OH community yet. Existing program analysis methods have several problems when handling the \arkts code.
To bridge the gap, we design and implement a framework named \toolname and make it publicly available as an open-source project. Our \toolname addresses the aforementioned problems and has already integrated a number of fundamental static analysis functions (e.g., control-flow graph constructions, call graph constructions, etc.) that are ready to be reused by developers to implement \OH App analyzers focusing on statically resolving dedicated issues such as performance bug detection, privacy leaks detection, compatibility issues detection, etc. Experiment results show that our \toolname achieves both high analyzing efficiency and high effectiveness. In addition, we open-sourced the dataset that has numerous real-world \arkts Apps.
\end{abstract}


\section{Introduction}

To support seamless interoperability among different devices, our community invents a new open-source mobile operating system called \OH, which is operated by the OpenAtom Foundation\cite{openatomcon} in China.
At the moment, the \OH ecosystem already has numerous applications~\cite{li2023software}. 
Considering that \OH is still in its early development stage, it shows great potential and promising future market prospects\cite{liu2024llm, chen2024model, sie2022analysis}.
However, as an independent all-scenario operating system, \OH features a brand-new application development paradigm and API that are not compatible with existing applications. 
Thus, a more user-friendly language, \arkts, has been introduced to \OH ecosystem.

In this context, we expect that the various issues (e.g., security, compatibility, performance, etc.)\cite{grech2015static, pan2020systematic, vassallo2020developers} that have been previously encountered by the Android and iOS ecosystems will not be less for \OH.
Hence, the various program analysis approaches proposed to address those issues will also need to be constructed for \OH.
Taking Android as an example, there are numerous advanced program analysis tools that safeguard the Android ecosystem, and the majority of these tools are based on the static analysis framework Soot\cite{vallee2010soot}.


\begin{figure}[h]
    \centering
    \includegraphics[width=\linewidth]{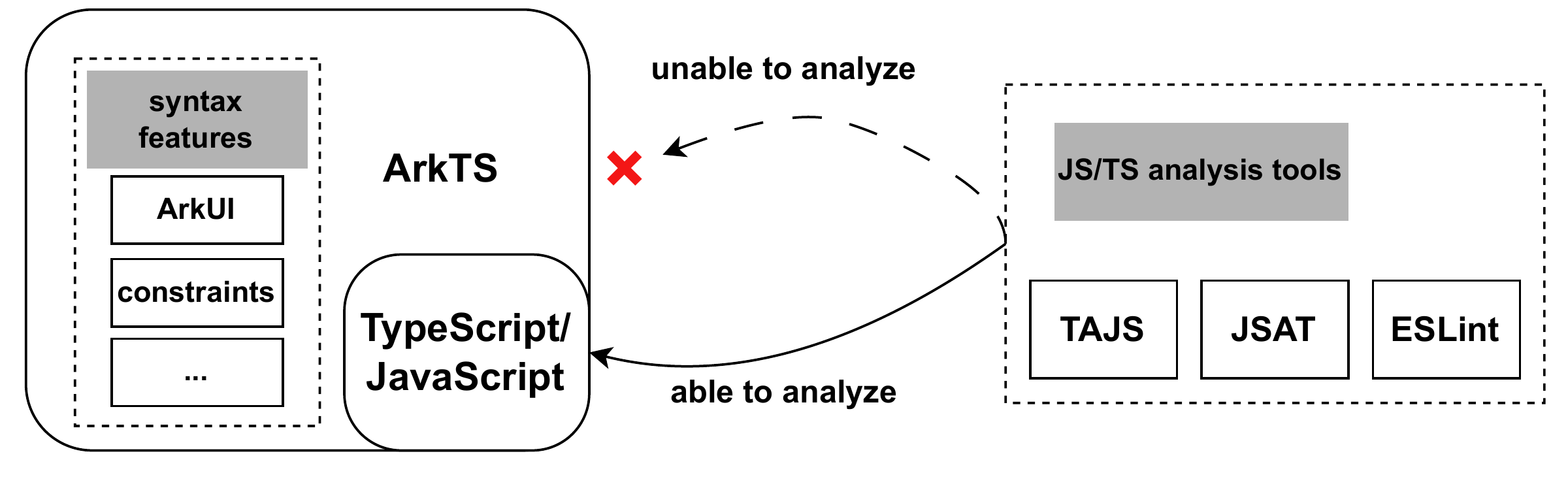}
    \caption{Existing JS/TS Analysis Tools Inapplicable to \arkts}
    \label{fig:background}
\end{figure}
Unfortunately, there is no such common static analysis framework available for the \OH community. As shown in Figure~\ref{fig:background}, although \arkts originates from TypeScript, it has introduced a multitude of innovative features, most notably the declarative UI and its accompanying new syntax. The significant differences between \arkts and TypeScript at the source code level lead to the addition of various new syntax nodes and structures in the Abstract Syntax Tree (AST) and will cause multiple kinds of errors in analyzing \arkts codes by existing methods. In addition, although ArkTS and TS have some similar features, we do not consider to translate the ArkTS codes to TS codes to enable the program analysis for ArkTS. The rule-based translation approaches are not stable, and the learning-based methods do not have enough training data. As an emerging programming language, ArkTS is in rapid development and will have more and more unique features. Therefore, we need to invent an independent static analysis framework to analyze the ArkTS codes.

To bridge the gap, we present to the community a prototype tool called \toolname \footnote{For the code related to \toolname, please refer to \url{https://gitee.com/OpenHarmony-sig/arkanalyzer}} by providing support for the unique features of \arkts in program analysis, which implements various software engineering approaches dedicated to scrutinizing \OH Apps.
The implementation of \toolname is adapted to the new features of the \OH system and the emerging \arkts language, achieving multi-dimensional analysis. Specifically, we propose our Code Representation module and Code Transformation module to address the definition mismatch and structure mismatch problems in existing program analysis methods. In addition, we collect the extra constraints of \arkts and solve them in our \toolname. The experiment results show that our method has high efficiency (within 10 seconds for analyzing a call graph of an App with thousands of lines of code) and effectiveness (93.75\% of accuracy in CHA and 87.95\% of accuracy in RTA).

The main contributions of our study are summarized as follows:
\begin{enumerate}[leftmargin=*]
    \item we provide a comprehensive analysis and empirical evaluations of \arkts, the new programming language for developing native applications on HarmonyOS, to identify the challenge in analyzing the \arkts by existing methods. 
    \item We propose and open-source a novel static analysis framework, \toolname, which addresses the problem of analyzing the \arkts program with unique syntax nodes and structures in AST. 
    \item We open-source a dataset of \arkts Apps to the community, which was collected from three official \OH repositories \footnote{The dataset is open-sourced on \url{https://bhpan.buaa.edu.cn/anyshare/zh-cn/link/AA5F769683A23C43B0A4D384D70C7505EB?_tb=none}}. We manually select high-quality Apps and repositories to improve the comprehensiveness and quality of evaluation results. 
    \item We conducted a comprehensive evaluation of \toolname. It demonstrates that the Intermediate Representation (IR) generated by our method is highly readable and our tool is efficient and effective.

\end{enumerate}


\section{Background: \arkts vs. TS}
\label{sec:backgroud}

In order to benefit from the existing ecosystem of TypeScript (TS), which has gained a large number of libraries, \arkts attempts to retain as many features as possible when extending the TypeScript language.
Nevertheless, in order to support a high-performance experience that is essential for Apps running on mobile devices, \arkts has to make some changes compared to its original design.
Specifically, there are two major kinds of unique features: (1) adding new features required by mobile Apps like ArkUI, and (2) constraining the flexibility of TypeScript, primarily its dynamic features that could impact execution performance.
We now detail these two types, respectively.

\subsubsection{ArkUI}

As a declarative UI framework, compared to the traditional procedural and imperative UI approaches, ArkUI focuses on the outcome of the UI description. It binds the UI to reactive data, which is more efficient as developers only need to concentrate on data management. Additionally, the declarative UI offers a declarative description akin to natural language, making it more intuitive. The industry has chosen declarative UI as the new generation model for application development and has undertaken a corresponding restructuring of the underlying UI component design to accommodate this paradigm shift.

\begin{figure}[h]
    \centering
    \includegraphics[width=0.9\linewidth]{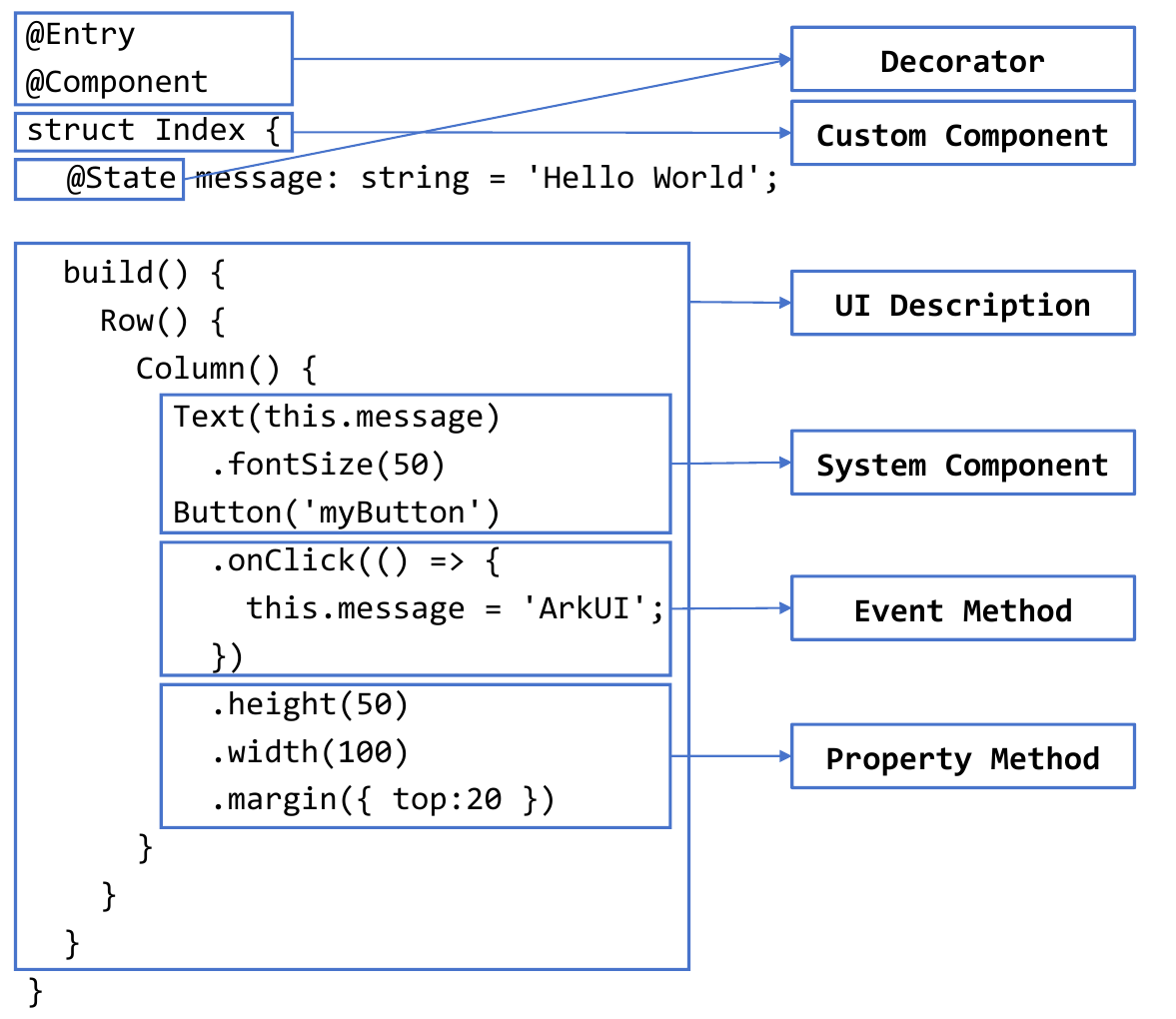}
    \caption{ArkUI Code Example.}
    \label{fig:ArkUI_example}
\end{figure}

As previously mentioned, the new syntactic structures introduced by ArkUI are one of the primary reasons why traditional JS/TS analysis tools cannot effectively analyze \arkts applications.
In order to help understand, we illustrate the components of ArkUI through a simple ArkUI code example, as shown in Figure~\ref{fig:ArkUI_example}.
\textbf{Decorator} features play a pivotal role, with elements like \texttt{@Component} marking custom components, \texttt{@Entry} specifying entry components, and \texttt{@State} indicating dynamic state variables that prompt UI updates upon modification. The \textbf{UI Description} is systematically defined within the \texttt{build()} method, detailing the UI's structural elements in a clear, declarative manner. \textbf{Custom Component} refers to reusable UI blocks, such as the Index structure, which can incorporate other elements and is designated by the \texttt{@Component} decorator. \textbf{System Component} includes fundamental and container components built into the framework, like \texttt{Column}, \texttt{Text}, \texttt{Divider}, and \texttt{Button}, offering readily accessible tools for developers. \textbf{Property Method} and \textbf{Event Method} allow for detailed customization and interaction handling within components; for instance, property methods like \texttt{fontSize()}, \texttt{width()}, \texttt{height()}, and \texttt{backgroundColor()} adjust visual aspects, while event methods such as \texttt{onClick()} facilitate user engagement strategies. This architecture not only simplifies the development process but also enhances the functionality and interactivity of the application interfaces. 

\subsubsection{Syntactic Constraints.} 
\label{bg:extra_constraints}
\arkts specification imposes constraints on overly flexible features in TypeScript that can affect development correctness or introduce unnecessary overhead during runtime. Even though these constraints may not cause TypeScript analysis tools to throw errors when analyzing \arkts code, as differences between the two languages, they are likely to affect the accuracy of the analysis results. For clearly presenting such Syntactic Constraints, here are some examples:
(1) \textbf{\texttt{any} Type}. \arkts mandates the use of static types to improve code clarity and performance. For example, it prohibits the use of the \texttt{any} type
, encouraging explicit type definitions that can be analyzed at compile time for correctness.
(2) \textbf{Object Layout.} \arkts does not allow changes to an object's structure
at runtime, such as adding or deleting properties, to optimize runtime performance and predictability.
(3) \textbf{Operator Semantics.} \arkts restricts certain operator semantics to encourage clearer code and avoid runtime overhead, such as disallowing the unary + operator on non-numeric types.
(4)\textbf{Structural Typing.} Unlike TypeScript, which supports structural typing
allowing objects with the same shape to be considered of the same type, \arkts requires explicit declarations, enhancing type safety and consistency.

\section{Preliminary Study}
\label{sec:arkts}

Recall that the \arkts language is extended from the widely used TypeScript (hereinafter referred to as TS) programming language.
When exploring the feasibility of static analysis for \arkts, we would like to first explore if \textbf{existing JS/TS static analysis tools can be directly applied to analyze \arkts code.}
If so, there is no need to specifically develop a static analysis framework for the \arkts language.



\subsection{JS/TS Analyzers Identification}

To delve into the aforementioned question, we first conduct an exploratory study to identify mainstream JS/TS static analysis tools.
We choose three tools that have been widely used in academic papers or industry: TAJS, JSAI, and ESLint.

\begin{itemize}[leftmargin=*]
    \item TAJS (Type Analysis for JavaScript)~\cite{jensen2009type} is a static program analysis tool designed to provide detailed and precise type information for JavaScript programs.
    TAJS not only detects common programming errors but also performs type inference and generates call graphs, among other analyses.

    \item JSAI~\cite{kashyap2014jsai} is another JavaScript static analysis platform that implements a range of analysis capabilities such as type inference, pointer analysis, control flow analysis, and constant propagation. 

    \item ESLint\cite{eslint}, a powerful and highly pluggable JavaScript code-checking tool, is currently one of the most widely used JS/TS code analysis tools. 
    ESLint supports modern JavaScript (ECMAScript) features and can integrate with various editors and build tools, thus enhancing development efficiency and code consistency. 
\end{itemize}


\subsection{Dataset}
\label{subsec:dataset}


To support the preliminary study, we need to form a real-world dataset. We collect Apps from three official \OH organization repositories: the \OH repository \cite{openharmonyProject}, \OH-SIG \cite{openharmonySig}, and \OH-TPC \cite{openharmonyTpc}. The main repository is the core codebase of the \OH project, containing the fundamental components of the operating system and serving as the primary interaction and contribution point for developers and contributors. The \OH-SIG repository supports specific interest groups (SIG), responsible for managing development in particular technical areas such as the graphics subsystem and the device driver subsystem. The \OH-TPC repository focuses on collecting and maintaining third-party open-source libraries, facilitating access for developers, and ensuring compliance with open-source standards.

It is important to note that the dataset used in this study does not include all applications, but rather underwent a selection process. We only select applications where the repository has more than 10 stars and the number of lines of \arkts code exceeds 100, ensuring that the dataset consists of applications with a certain level of quality. 
As of April 10, 2024, the collected dataset includes 371 \OH repositories, 100 \OH-SIG repositories, and 147 \OH-TPC repositories. Ultimately, we collected 618 \OH applications.



\subsection{Results}
\label{pre_results}

Our results indicate that the existing tools (\ie TAJS, JSAI, and ESLint ) are entirely incapable of analyzing \arkts applications comprehensively. Among the three tools we tested, none were able to fully analyze any of the applications without encountering errors.

Upon reviewing the specific error descriptions, we observed that existing tools might be successful in analyzing code that does not deviate from standard TypeScript. However, when attempting to analyze code that incorporates new features unique to \arkts, such as ArkUI and extra constraints, the tools produced "Parsing error" messages. Our collected applications have 7199 \arkts code files. 3601 and 3138 files cannot be analyzed by \textbf{TAJS} and \textbf{ESLint} due to the ArkUI-related problems, respectively. 3585 and 2914 files cannot be analyzed by \textbf{TAJS} and \textbf{ESLint} due to the extra constraint problems, respectively. For \textbf{JSAI}, it even cannot successfully analyze any of the \arkts code files, and its identical error messages prevent us from classifying the causes of the errors.
Indeed, static analysis techniques are usually sensitive to programming languages, as different languages have different syntax rules, different semantics, and different language features.

\vspace{-0.2cm}
\findingbox{Finding of the Preliminary Study}{Existing JS/TS-based static analyzers cannot be applied to analyze \OH Apps. There is hence a strong need to design and implement dedicated static analyzers for \OH.}

\section{Methodology}

In this section, we will detail our solution \toolname.

\subsection{Motivation}

According to the results in Section~\ref{pre_results}, we need to invent new methods to address the problems in analyzing the \arkts using existing analyzers. Before presenting our method, we first analyze the errors in existing analyzers:

\begin{itemize}[leftmargin=*]

\item {\small\textbf{Failure by ArkUI - Definition Mismatch:}} As a new programming language, \arkts defines new declaration keywords such as \texttt{struct} with a unique internal structure. Those structures do not exist in TypeScript, so analysis tools cannot recognize them. In addition, ArkUI allows using a number of decorators without prior definition (\eg \texttt{@Entry}, \texttt{@Component}, \texttt{@State} in Figure~\ref{fig:ArkUI_example}), which is not permitted in TypeScript and would cause errors in the existing compilers and analysis tools. We call this kind of error Definition Mismatch. \textbf{To address it, in the Code Representation module of our approach, \toolname enable the modeling of mismatched definitions within \arkts using a newly designed AST.}

\item {\small\textbf{Failure by ArkUI - Structure Mismatch:}}  ArkUI contains nested system components (\eg \texttt{Row()} and \texttt{Column()} in Figure~\ref{fig:ArkUI_example}), which are used in a way similar to the structure of function declarations, but without the `function' keyword, and allow continuous built-in function calls at the end of the component. Therefore, the function structures of some \arkts programs are not compatible with the existing analyzers which are designed for analyzing TypeScript's syntax rules. We call this kind of error Structure Mismatch. \textbf{To this end, we propose a Code Transformation module to simplify the \arkts code and support the analysis of the mismatched function structures.}

\item {\small\textbf{Extra Constraints:}} As we mentioned in the Background, compared with TypeScript, \arkts uses extra constraints to optimize the development correctness and runtime overhead. These extra constraints such as Any Type, Object Layout, Operator Semantics, and Structural Typing (see Section~\ref{bg:extra_constraints}) are not compatible with existing program analysis methods for TypeScript. \textbf{Because we fix the extra constraint errors in an ad-hoc manner, we omit them in this paper.}

\end{itemize}

Therefore, we design and implement in this study a prototype tool called \toolname, which aims at bridging the gap between the existing program analysis methods and \arkts. Our method can remove the aforementioned errors while having high analyzing efficiency for \arkts.

\subsection{Overview of \toolname}
\label{sec:design}

\begin{figure}[h!]
    \centering
    \includegraphics[width=\linewidth]{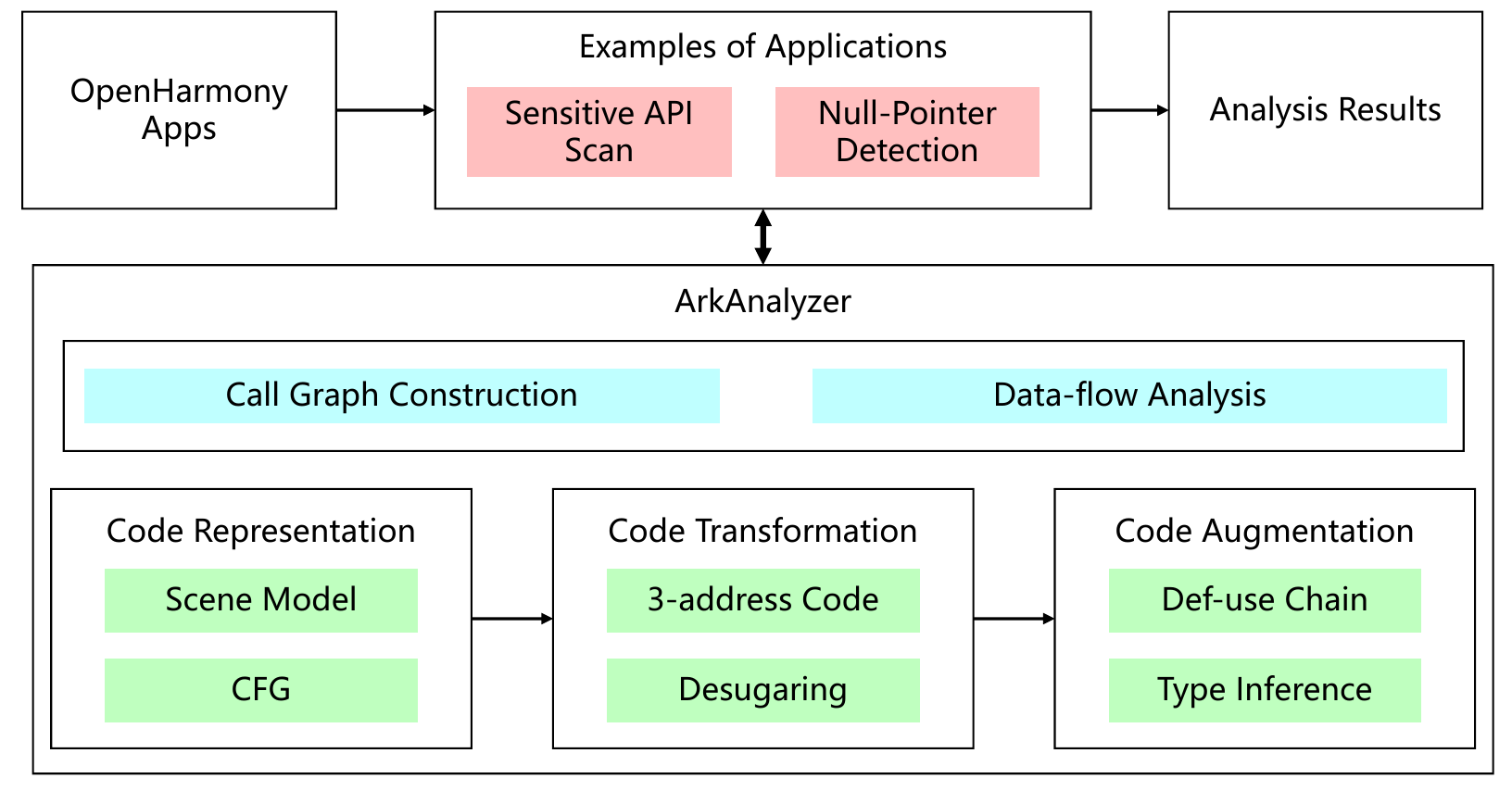}
    \caption{Overview of the Design of \toolname.}
    \label{fig:overview}
\end{figure}

We now briefly introduce the core functions included in the \toolname framework.
As shown in Fig.~\ref{fig:overview}, \toolname by itself is a framework dedicated to facilitating the implementation of App analyzers such as tools for detecting the usages of sensitive APIs or characterizing Null-pointer issues.
Inside \toolname, the input App code will be handled in two layers, with the bottom layer responsible for basic analyses and the upper layer for more advanced analyses.

Specifically, in the bottom layer, \toolname starts with the AST generated by the ArkTS compiler to model the application source code, and then transforms and augments the code to facilitate subsequent analysis.
Then, in the upper layer, \toolname leverages the outputs of the first layer to represent the App code with more advanced data structures (such as call graphs and inter-procedural data flows). 


We now detail these modules to help readers better understand the design of \toolname.

\subsection{Code Representation}
\label{subsec:scene}

\begin{figure}[h]
    \centering
    \includegraphics[width=\linewidth]{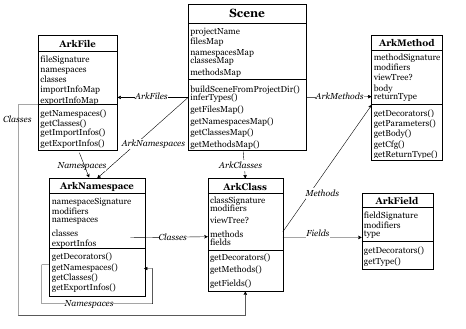}
    \caption{The design of the core classes designed for representing code under analysis.}
    \label{fig:scene}
\end{figure}

In \toolname, We employ a newly designed AST specifically tailored for \arkts analysis. 
This AST is a product of the \arkts compiler, designed to accommodate the new features of \arkts and support the modeling of ArkUI code segments.
In each analysis, \textbf{Scene} serves as the entry point and contains comprehensive information about the project. It is designed to provide a unified context environment, enabling access to and manipulation of various program details during the analysis process. 
Figure~\ref{fig:scene} illustrates the core classes managed by the \textbf{Scene} model. We now detail the representative ones.

\textbf{ArkFile} represents each individual file, simplifying the management of project files. 
In the context of the \arkts language, \textbf{ArkNamespace} object is designed to encapsulate the information and structure within a namespace. This facilitates access to and handling of classes and methods within the namespace scope, maintaining the logical organizational structure of the code.
Given that \arkts supports object-oriented programming, the analysis of object-oriented structures is essential. \textbf{ArkClass} object represents a class in the object-oriented paradigm, encapsulating internal structural information such as attributes and methods.
Methods and Fields of a class are abstracted into \textbf{ArkMethod} and \textbf{ArkField} classes, respectively.

To address the definition mismatch problem, we abstract \texttt{struct} as ArkClass because it encompasses its own properties and functions, bearing similarities to \texttt{class} in structure. The ArkClass corresponding to \texttt{struct} will have specific identifiers and special properties, such as \texttt{viewTree}, which represents its corresponding ArkUI component tree. Through the component tree, it is possible to deduce which components the \texttt{struct} uses and the composition relationships between them.
Additionally, we have introduced an abstract class \textbf{Decorators} to correspond to the extensive use of decorators in ArkUI. Each namespace, class, method, and field can obtain their corresponding decorators through specified interfaces.


\begin{figure}[h]
    \centering
    \includegraphics[width=0.8\linewidth]{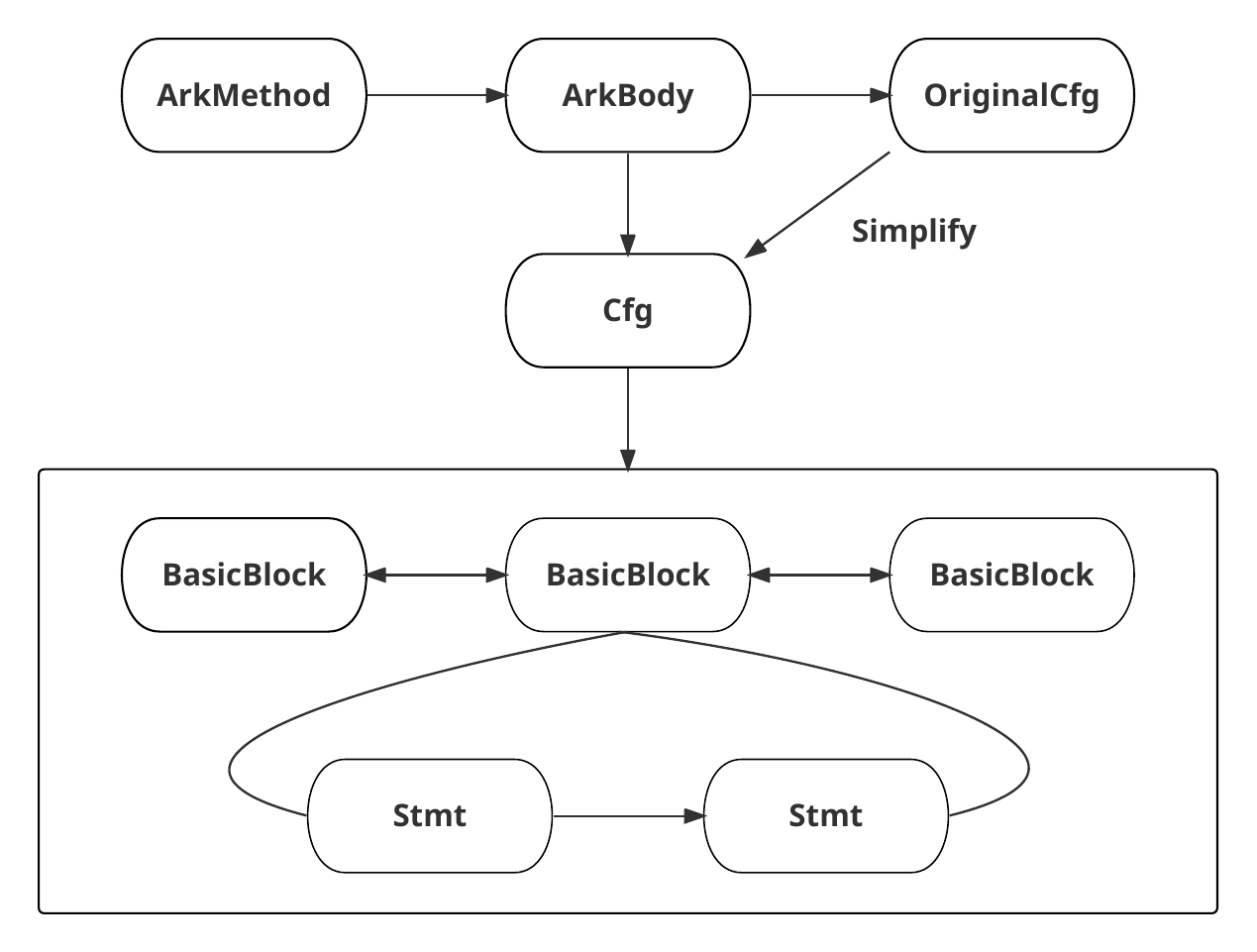}
    \caption{The design of the ArkMethod class.}
    \label{fig:arkbody}
\end{figure}

As shown in Figure~\ref{fig:arkbody}, 
the actual code of a given method (i.e., \textbf{ArkMethod}) will be recorded in a so-called \textbf{ArkBody} class, which is further represented via two Control Flow Graphs, namely \textbf{OriginalCfg} and \textbf{Cfg}. \textbf{Cfg} is the simplified version of the \textbf{OriginalCfg}, which represents the control-flow graph built based on the original code of the method.
Each \textbf{Cfg} is composed of several \textbf{BasicBlock}, and each \textbf{BasicBlock} contains a series of sequentially executed lines of code (i.e., without branches).
In this work, each line of code is recorded via a \textbf{Stmt} class.





\subsection{Code Transformation}
\label{subsec:transformation}

After code representation, we leverage a code transformation step to mitigate the structure mismatch problem that may cause difficulties when performing existing analyzers to \arkts.
The \arkts compiler directly transforms the source code into bytecode\cite{arkcomplier}. Although intermediate code, such as Panda IR, can be obtained through disassembly tools, it is more bytecode-oriented and lacks readability, which contradicts \toolname's design philosophy of high readability and user-friendliness. Therefore, we need to design our own form of IR and establish the corresponding transformation rules.

Specifically, we take two approaches to transform the \arkts code:
(1) Change the code to align with the three-address form, and (2) Transform the code to mitigate certain features such as removing loops, naming anonymous classes or functions, transforming system components into regular code form, etc.
We now detail these two approaches, respectively. Thus, our simplified code can be handled by the following analyzing steps (\ie Section~\ref{subsec:augmentation}, \ref{subsec:cg}).

\subsubsection{Three-address Code.}
Three-address code is a common intermediate code representation format, for which each line of code is ensured to have at most three operands (addresses). In addition to basic arithmetic expressions, syntactic constructs such as object property access, function calls, and array indexing also need to be converted into three-address code. Converting source code to three-address code has significant benefits for program analysis. 
Indeed, Its simple and uniform format simplifies the program structure, making it easier to handle and analyze.

In \toolname, some of the representative conversion rules transforms source code to three-address format are highlighted in Table~\ref{lab:rewriting_rules}. Given a complex statement, we will divide it into several simple statements.
To do so, we will create temporary variables to bridge these simple statements (i.e., keep the same code semantics).
Regarding complex function call expressions (including expressions as arguments, nested calls, and call chains), \toolname will progressively break them down in the order of execution. The result of each step is stored in temporary variables. By applying these rules, complex code is transformed into a more manageable three-address code format, laying the groundwork for further optimization and analysis.

\begin{table}
  \caption{Rules applied to achieve three-address code.}
  \label{lab:rewriting_rules}
  \begin{center}
  \resizebox{1\linewidth}{!}{
    \begin{tabular}{ c |c|c |c} 
     \toprule
          No & Rule & Before & After \\\hline
1  & Complex Expression & \makecell[l]{\textit{x = a.b + c.d}} & \makecell[l]{\textit{temp1 = a.b}\\ \textit{temp2 = c.d}\\ \textit{x = temp1 + temp2}} \\ \hline
2  & Expression Parameter & \makecell[l]{\textit{x = fun(a + b)}} & \makecell[l]{\textit{temp1 = a + b}\\ \textit{x = fun(temp1)}} \\ \hline
3  & Nested Function Call & \makecell[l]{\textit{x = funA(funB())}} & \makecell[l]{\textit{\_ret = funB()}\\ \textit{x = funA(\_ret)}} \\ \hline
4  & Subscript Operation & \makecell[l]{\textit{x = funA()[1]}} & \makecell[l]{\textit{\_ret = funA()}\\ \textit{x = \_ret[1]}} \\ \hline
5  & Call Chain Splitting & \makecell[l]{\textit{f1().f2().f3()}} & \makecell[l]{\textit{x = f1()}\\ \textit{y = x.f2()}\\ \textit{z = y.f3()}} \\
     \bottomrule
    \end{tabular}
 }
    \end{center}
\end{table}

\subsubsection{Code Desugaring}

After representing the code to three-address format, we go one step deeper to further simplify the code by conducting a desugaring phase.
Syntactic sugar refers to the addition of certain syntax features in a programming language that make the code more concise and readable without changing the language's functionality. These features typically simplify common programming patterns, allowing programmers to express the same logic in a more straightforward manner.
However, those syntactic sugars, although being more friendly to developers, do make code analyses more complex.
To that end, towards preventing syntactic sugar from hindering code analysis, we perform a code desugaring phase by transforming code using syntactic sugar into a semantically equivalent form. 

Table~\ref{lab:desugaring_rules} highlights some of the representative transformation rules adopted by \toolname.
First, increment operators (like \texttt{i++}) and compound assignments (like \texttt{i /= 5}) need to be converted to standard assignment operations.
Template strings, using string interpolation, should be transformed into string concatenation operations. 
Arrow functions and Anonymous functions should be converted into regular function expressions. 
Object literals should be converted into explicit class definitions and instantiations. 
For better supporting the representation of control flows, we also take the opportunity to simplify the code by transforming \texttt{if-else} and loop statements into structures with explicit labels and jumps. These transformations standardize the code, making it easier for subsequent analyses.

The ninth row in the table demonstrates \toolname's handling of nested system component code within ArkUI, which can cause the structure mismatch problem in existing analyzers.\toolname maps each system component to the corresponding interface in the \OH SDK. First, each component is associated with its corresponding \texttt{create} function and \texttt{pop} function. Then, subsequent function calls on the component are applied to the temporary variable returned by the \texttt{create} function. In this way, the special function in ArkUI (the \texttt{build} function of the \texttt{struct}) is transformed into a regular code format, resolving the structure mismatch issue.

\begin{table}
  \caption{Transformation rules applied to simplify code at the IR level.}
  \label{lab:desugaring_rules}
  \begin{center}
 \resizebox{1\linewidth}{!}{
    \begin{tabular}{ c |c|c |c} 
     \toprule
No  & Rule & Before & After \\\hline
1   &  Increment Operators & \makecell[l]{\textit{i++}} & \makecell[l]{\textit{i = i + 1}} \\ \hline
2   &  Compound Assignment & \makecell[l]{\textit{i /= 5}} & \makecell[l]{\textit{i = i / 5}} \\ \hline
3   & Template Strings & \makecell[l]{\textit{greet = !\$\{name\}!}} & \makecell[l]{\textit{temp1 = name + `!'}\\ \textit{greet = `!' + temp1}} \\ \hline
4   & Arrow Function & \makecell[l]{\textit{fun = (x) => x + 1}} & \makecell[l]{\textit{def Anonymous\_1(x)}\\ \textit{~~~~return x + 1}\\ \textit{fun = Anonymous\_1}} \\ \hline
5 & Anonymous Function & \makecell[l]{\textit{set(fun() \{} \\ \textit{~~~~...} \\ \textit{\}, 1)} } & \makecell[l]{\textit{def Anonymous\_1():}\\ \textit{~~~~...}\\ \textit{set(Anonymous\_1, 1)} } \\ \hline
6 & Anonymous Class & \makecell[l]{\textit{let x = \{name: `a'\};} } & \makecell[l]{\textit{class Anonymous\_1\{}\\ \textit{~~~~name: string} \\ \textit{\}} \\ \textit{x = new Anonymous\_1()} \\ \textit{x.name = `a'} } \\ \hline
7 & Control Flow (if) & \makecell[l]{\textit{if (x > 0)} \\ \textit{~~~~x++} \\ \textit{else} \\ \textit{~~~~x-{}-}} & \makecell[l]{\textit{label1 :} \\ \textit{~~~~if (x > 0)} \\ \textit{~~~~~~~~goto label2 label3} \\ \textit{label2:} \\ \textit{~~~~x++} \\ \textit{~~~~goto label4} \\ \textit{label3:} \\ \textit{~~~~x-{}-} \\ \textit{~~~~goto label4} \\ \textit{label4:} \\ \textit{  //following statements}}  \\ \hline
8 & Control Flow (while) & \makecell[l]{\textit{while (x > 0)} \\ \textit{~~~~x-{}-} \\ \textit{console.log(x)} } & \makecell[l]{\textit{label1 :} \\ \textit{~~~~if (x > 0)} \\ \textit{~~~~~~~~goto label2 label3} \\ \textit{label2:} \\ \textit{~~~~x-{}-} \\ \textit{~~~~goto label1} \\ \textit{label3:} \\ \textit{~~~~console.log(x)}} \\ \hline
9 & System Component & \makecell[l]{\textit{Row ()\{} \\ \textit{~~~~Column ()\{} \\ \textit{~~~~\}.height(100)} \\ \textit{\}} } & \makecell[l]{\textit{temp0 = RowInterface.create()} \\ \textit{temp1 = ColumnInterface.create()} \\ \textit{temp1.height(100)} \\ \textit{ColumnInterface.pop()} \\ \textit{RowInterface.pop()} } \\ \bottomrule

    \end{tabular}}
 
    \end{center}
\end{table}

\subsection{Code Augmentation}
\label{subsec:augmentation}

The code representation and transformation steps have greatly reduced the complexity of the code under analysis.
However, there is still common information that is constantly required by follow-up analyzers but is not yet available in the current code representation.
To further facilitate the implementation of App analyzers, we add another step to \toolname to further augment the code.
Specifically, we pre-calculate data-flow information for each method by building a def-use chain and the type information for local variables based on a set of pre-defined rules (more advanced.\footnote{At this stage, only lightweight analyses are considered for the sake of performance, \ie data-flow analysis is limited within methods, type analysis is implemented without leveraging points-to analysis. More advanced analyses are also supported by \toolname but are at later stages.})
We now detail these two sub-steps, respectively.

\subsubsection{Def-use Chain}
\label{subsubsec:data_flow}

Data flow analysis is used to track the path from the definition of a variable to its usage within a program. The primary technique frequently adopted by our community to record such data dependencies in the program is to build the so-called def-use chains.
The chains are considered important for optimizing compilers, code refactoring, detecting potential errors, and identifying vulnerabilities. Analyzing these chains within a single program or function can help understand how local variables are initialized and utilized, thereby ensuring the correctness and efficiency of data flow.


\subsubsection{Type Inference}
\label{susubbsec:type_infer}

Compared to TS, \arkts imposes stricter type restrictions but still supports implicit type declarations. \toolname has formulated a series of rules for analyzing code statements to infer the type information of variables and other syntactic elements in the code\cite{10.1145/3510003.3510061}.
\toolname conducts a comprehensive scan of code statements within a project, initially extracting type information from individual statements and assigning it to corresponding variables. If direct inference is not possible, type propagation will be carried out based on contextual information. 

The specific rules leveraged in this step are listed in Table~\ref{lab:type_inference_rules}. Calculations and comparisons between simpler primitive types can directly determine the result's type. We also determine the declared class based on the literal following the $new$ keyword. Additionally, by referencing the declarations of corresponding classes, methods, and properties, we parse the types of the respective components within the statement.

\begin{table}
  \caption{Rules applied to infer types.}
  \label{lab:type_inference_rules}
  \begin{center}
 \resizebox{1\linewidth}{!}{
    \begin{tabular}{c|c|p{0.5\linewidth}|c} 
\toprule
\textbf{No}  & \textbf{Rule} & \textbf{Pattern}& \textbf{Type of x}\\\hline
1   & Compare& x = a op b,
op $\in$ {Eq, NotEq, Lt, LtE, Gt, GtE, Is, IsNot, In, NotIn}& bool\\ \hline
2   & BinOp& x = string * number
x = bool * number& str
bool\\ \hline
3   & Heap Object Create& x = new ClassA()& ClassA\\ \hline
 4& Return& x = func()&func's return type\\ \hline
 5& Field Reference& x = ClassA.field&ClassA.field type\\ 
 \bottomrule    
    \end{tabular}}
    \end{center}
\end{table}

\subsection{Call Graph Construction}
\label{subsec:cg}

Call graph is a fundamental data structure that is required by many analysis tasks and is essential to support project-wide analyses.
Call graph generally represents the relationships between method invocations within the program. In a given call graph, nodes represent methods, and directed edges signify the calling relationships initiated by the caller method pointed to the callee method. In this section, we will discuss the core algorithms adopted by \toolname for call graph construction. We have implemented two algorithms: (1) Class Hierarchy Analysis (CHA) and (2) Rapid Type Analysis (RTA)\cite{10.1145/506315.506316}, for which we will detail them respectively in this section.

\subsubsection{CHA: Class Hierarchy Analysis}
\label{subsec:cha}

\arkts is an object-oriented programming language. To support the object-oriented features, \toolname organizes the project under analysis in the form of classes, and their inheritance relationships are recorded, which is referred to as a class hierarchy tree.

The CHA algorithm builds the call graph by parsing the invocation statements within the code to identify basic invocation relationships and form the call graph, e.g., building an edge from method \texttt{$m_1$} to method \texttt{$m_2$}.
It then augments the call graph by adding new edges based on the aforementioned hierarchy tree.

\begin{lstlisting}[caption=Example code snippet for demonstrating the principles of constructing call graphs., label=lst:cg_code]
function makeAnimalSound(animal: Animal) {
    animal.sound();
}

function main() {
    let dog = new Dog();
    let cat = new Cat();
    makeAnimalSound(dog);
}
\end{lstlisting}

Taking Listing \ref{lst:cg_code} as an example, which illustrates a code snippet (omitting related class definitions), and the actual class hierarchy is presented in Figure \ref{fig:class_hierarchy_in_example}, where all classes have \texttt{sound} method.
In this case, when \texttt{Animal.sound} is called, \toolname will add three new edges (i.e., \texttt{$makeAnimalSound \to Dog.sound$}, \texttt{$makeAnimalSound \to Cat.sound$}) and \texttt{$makeAnimalSound \to Cow.sound$}) to the call graph in \ref{fig:CHA} because these three edges could also be true due to the polymorphic characteristic, one of the core features adopted in the object-oriented concept. 
Observant readers may have already noticed that the CHA algorithm offers a call graph that is as comprehensive as possible, attempting to record all the possible calling relationships.
This, however, will unavoidably introduce incorrect edges that subsequently would lead to false positive results for downstream analyzers.

\subsubsection{RTA: Rapid Type Analysis}

\begin{figure*}[ht]
    \centering
    \begin{subfigure}{0.23\textwidth}
        \centering
        \includegraphics[width=\linewidth]{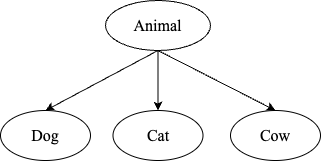}
        \caption{Part of Class Hierarchy}
        \label{fig:class_hierarchy_in_example}
    \end{subfigure}
    \hspace{10pt}
    \begin{subfigure}{0.27\textwidth}
        \centering
        \includegraphics[width=\linewidth]{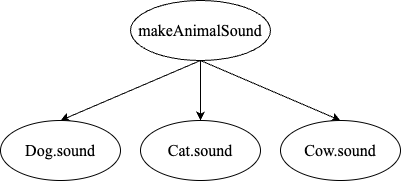}
        \caption{CG By CHA}
        \label{fig:CHA}
    \end{subfigure}
    \hspace{10pt}
    \begin{subfigure}{0.2\textwidth}
        \centering
        \includegraphics[width=\linewidth]{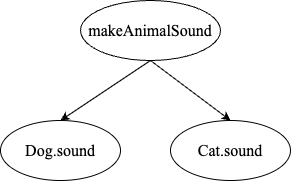}
        \caption{CG By RTA}
        \label{fig:RTA}
    \end{subfigure}
    
    \caption{Call graph construction.}
    \label{fig:code_example}
\end{figure*}

Building on the CHA, the RTA algorithm imposes certain constraints to filter potential calling relationships, thereby reducing the over-approximations inherent in CHA. 
During the construction of RTA, the actual creation of heap objects (i.e., instances created via the \texttt{new} keyword) is tracked and recorded. Upon encountering a method call statement and identifying potential call targets from the class hierarchy, RTA uses whether the class of the call target has been modeled as a criterion. It removes all methods from the call target that have not been modeled, ensuring that only relevant methods are considered. 
Given the same code snippets shown in Listing~\ref{lst:cg_code}, since only classes \texttt{Dog} and \texttt{Cat} are instantiated (i.e., class \texttt{Cow} is not instantiated), the edge \texttt{$makeAnimalSound \to Cow.sound$} will be removed by the RTA algorithm(cf. Figure~\ref{fig:RTA}), resulting in preciser call graph compared to that built by the CHA algorithm.

\section{Evaluation}
\label{sec:demo}

We evaluate the efficiency and accuracy of \toolname while exploring the readability of the IR\footnote{Readability of static analyzer's IR is considered very important as developers often need to read the IR to debug the analyzer (e.g., to understand its behavior).} designed by \toolname and the capability of supporting the implementation of advanced analyzers. Here are the details of the dataset and experiment environment:


\paragraph{Dataset} Recall that we have formed a dataset of 618 Apps when performing the preliminary study, as discussed in Section~\ref{subsec:dataset}. In that dataset, we have endeavored to collect and select all the high-quality \OH Apps that are available to the public. In this section, we reuse this dataset for evaluation.

\paragraph{Experiment Environment}
Our method is evaluated on a workstation with Intel(R) Core(TM) i7-14700KF CPU, 16GB of RAM and the 64-bit Windows 11 OS.


\subsection{Efficiency of \toolname}
\label{sec:efficiency}

The performance of static code analysis tools is crucial, as they must provide feedback as quickly as possible while maintaining analytical accuracy\cite{9400688}, especially in Continuous Integration/Continuous Delivery (CI/CD) pipelines. 
As previously mentioned, the Scene is the core structure of \toolname, and all analyses depend on the construction of the Scene. Therefore, to evaluate its performance, we tested the time required to construct Scenes for all applications in the dataset. Additionally, we measured the time taken for call graph analysis to demonstrate the high performance of \toolname.

\begin{figure*}[ht]
    \begin{minipage}{0.32\textwidth}
        \centering
        \begin{subfigure}{\linewidth}
            \includegraphics[width=\linewidth]{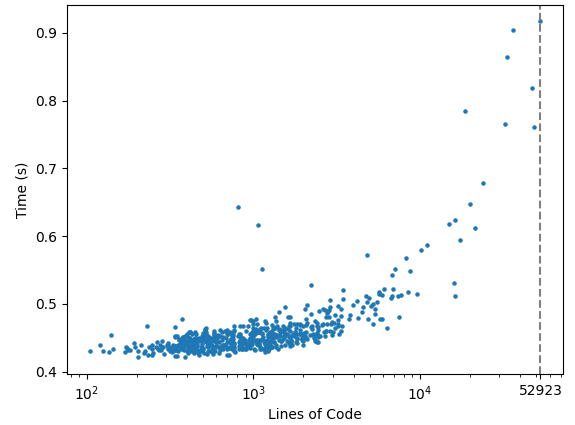}
            \caption{Time of Scene Construction}
            \label{fig:sandianscene}
        \end{subfigure}
    \end{minipage}
    \hfill
    \begin{minipage}{0.32\textwidth}
        \centering
        \begin{subfigure}{\linewidth}
            \includegraphics[width=\linewidth]{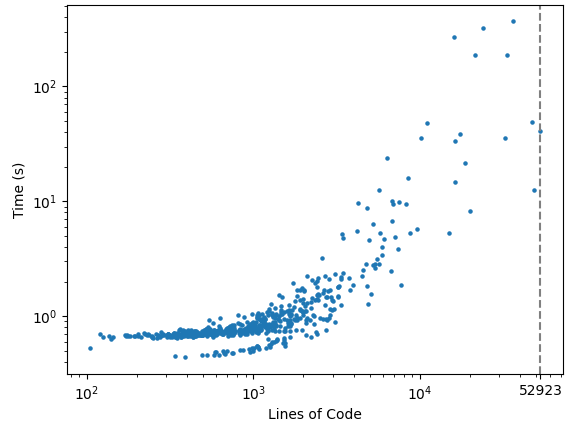}
            \caption{Time of Scene and CHA Construction}
            \label{fig:sandianCHA}
        \end{subfigure}
    \end{minipage}
    \hfill
    \begin{minipage}{0.32\textwidth}
        \centering
        \begin{subfigure}{\linewidth}
            \includegraphics[width=\linewidth]{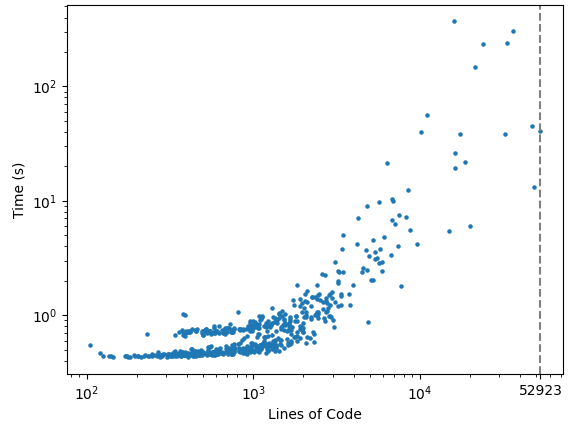}
            \caption{Time of Scene and RTA Construction}
            \label{fig:sandianRTA}
        \end{subfigure}
    \end{minipage}
    \caption{\toolname Analysis Performance Scatter Plot}
\end{figure*}

Figure~\ref{fig:sandianscene} shows the result of scene build time distribution. The majority of scene build times are concentrated between 0.4 and 0.5 seconds. A small portion of more complex scenes take longer to build, but all are built within 1 second.
Figure~\ref{fig:sandianCHA} and Figure~\ref{fig:sandianRTA} demonstrate the performance of \toolname during CHA and RTA analyses. Due to significant variations in some data, we applied a logarithmic transformation to both the x and y axes to enhance the clarity of the pictures. It turns out that whether using CHA or RTA, 
analysis of applications within a thousand of lines of code can be completed within 1 second, and analysis of applications with thousands of lines of code can be completed within 10 seconds. This result demonstrates the efficiency of call graph analysis.




These experimental results demonstrate that \toolname exhibits excellent performance in application analysis, with extremely high code processing efficiency, fully reflecting its effectiveness and stability in handling applications of varying scales.

\subsection{Accuracy of \toolname}
\label{sec:accuracy}

For the sake of simplicity, we validate the overall accuracy of \toolname by testing the accuracy of the call graph module, which is considered one of the most crucial feature for program analysis tools.

\begin{table}[ht]
\centering
\caption{the accuracy of \toolname in analyzing call graph}
\label{tab:cgAccuracy}
\begin{tabular}{c|c|c|c|c|c}
\toprule
\textbf{Dataset} &  \textbf{Algorithm} & \textbf{TP} & \textbf{All} & \textbf{Precision} & \textbf{Recall}  \\ \hline
\multirow{2}{*}{Benchmark}  & CHA & 80 & 80 & 96.39\% & 100\% \\ \cline{2-6} 
  & RTA & 78 & 80 & 100\% & 97.50\%  \\ \hline
\multirow{2}{*}{Real Apps}  & CHA  & 351 & 375 & 99.72\% & 93.75\%  \\\cline{2-6} 
  & RTA  & 332 & 375 & 99.70\% & 87.95\%\\ \bottomrule
\end{tabular}
\end{table}

We conducted experiments on two datasets: one consisting of a series of benchmark test sets that we specified, and the other comprising randomly selected samples from the dataset of real HarmonyOS applications mentioned in Section~\ref{subsec:dataset}. 
For each dataset, we performed call graph analysis using both CHA and RTA. The tests were conducted at the level of call chains, where we compared the call chains obtained by \toolname with those manually verified, calculating precision and recall. 
The results are shown in Table~\ref{tab:cgAccuracy}.

For the Benchmark dataset, the CHA algorithm achieved a precision of 96.39\% and a recall of 100\%, correctly identifying all 80 true positives (TP) out of 80 total calls. On the same dataset, the RTA algorithm yielded a precision of 100\% and a recall of 97.50\%, with 78 true positives correctly identified out of 80 calls. The CHA's strategy is to consider all methods with the same name in subclasses as potential call targets when the invoked method is identified within a calling statement and when the calling object has subclasses. This approach results in false positives in benchmark testing sets for the CHA algorithm, and RTA apply a stricter type check to avoid false positives.
For the Real Apps dataset, CHA achieved a precision of 99.72\% and a recall of 93.75\%, identifying 351 true positives out of 375 total calls. The RTA algorithm on this dataset produced a precision of 99.70\% but a slightly lower recall of 87.95\%, correctly identifying 332 true positives out of 375 calls. Both CHA and RTA achieved high precision. However, in certain complex invocations or specific method calls(function pointers and rare instances of HarmonyOS SDK calls), the algorithm may fail to accurately locate method declarations, resulting in false negatives. Furthermore, the type checking employed by the RTA can lead to the incorrect exclusion of certain method calls.
Overall, the results indicate that both algorithms performed well, with CHA showing slightly higher precision and recall results than RTA, particularly on the real applications dataset.


\subsection{Readability of \toolname-IR}

To evaluate the readability of intermediate representation (IR) in \toolname, we designed and implemented a questionnaire survey. We searched for participants based on their expertise in programming and ArkTS. The 17 participants' programming experience ranges from one year to eight years. Among them, six participants are experts of ArkTS while others only have few ArkTS skills.
The questionnaire included six rating items focused on combination operations, conditional branches, arrays and loops, function calls, anonymous functions, and a composite score. The rating scale employed a five-point system, where 1 indicated ``very difficult to understand" and 5 indicated ``very easy to understand." Additionally, the questionnaire featured a non-mandatory open-ended question to gather specific feedback from respondents on challenging aspects of the intermediate code. A total of 17 valid responses were collected, and the average scores for the rating items are presented in the Table~\ref{tab:survey}.

\begin{table}[h]
    \centering
    \caption{Average Scores for Readability of Intermediate Code. 
    }
    \label{tab:survey}
    \begin{tabular}{c|c|c|c|c}
        \toprule
        Item               & Average & Median & Min & Max \\ \hline
        Combination Operations & 4.82  & 5 & 4 & 5        \\ 
        Conditional Branches   & 4.37  & 5 & 1 & 5          \\
        Arrays and Loops       & 3.71  & 4 & 2 & 5          \\
        Function Calls         & 3.65  & 4 & 1 & 5          \\
        Anonymous Functions     & 4.00  & 5 & 2 & 5          \\
        Composite Score        & 4.06  & 4 & 3 & 5          \\ \bottomrule
    \end{tabular}
\end{table}

The results reveal differences in readability among various types of intermediate code. Specifically, combination operations received the highest average score, indicating their clear structure and ease of understanding. In contrast, arrays and loops, as well as function calls, had lower scores, respectively, likely due to their complexity and the inclusion of extraneous and redundant information, as further confirmed by the open-ended responses. Both the average and median of the composite scores reached 4, suggesting that most intermediate code performs well in terms of readability. 

\subsection{Capability of \toolname}

To demonstrate the the practical utility of \toolname, we now present two concrete App analyzers that are implemented on top of \toolname.
These two examples are selected because of their simplicity (with only a few lines of code). The \toolname by itself is designed to be as generic as possible and thereby it should be able to support the implementation of as many App analyzers (including ones involving complicated logic) as possible.

\subsubsection{Sensitive API Scan}

Scanning sensitive API in the code is crucial for ensuring the security, performance, and privacy of software. Sensitive APIs may involve accessing personal information or system-level resources of users\cite{asemi2023study}. 

\toolname enables precise and convenient API scanning. As previously mentioned, 
\toolname provides various call graph analysis algorithms that allow developers to accurately identify specific functions in projects even with extensive usage of advanced object-oriented features such as inheritance and polymorphism.

Listing~\ref{code:scan} provides an example of scanning code for locating log invocations. Given a \texttt{scene} and an array of \texttt{MethodSignature} as entry points, we can obtain the corresponding project call graph. 
The returned result is a map, with the key being the caller and the value being the callee. By traversing the map, it is very easy to find out which functions call the target function. 

\begin{lstlisting}[caption={Code snippet to locate the usage of a given API. }, label={code:scan}]
function scanLog(scene:Scene, entryPoints: MethodSignature[], targetMethodSig: MethodSignature) {
    let callGraph = scene.makeCallGraphCHA(entryPoints);
    let calls = callGraph.getDynEdges();
    
    calls.forEach((callees: Set<MethodSignature>, caller: MethodSignature) => {
        if (callees.has(targetMethodSig)) {
            console.log(caller.toString());
        }
    });
}
\end{lstlisting}

This example demonstrates the usefulness of \toolname, i.e., with the help of \toolname, one only needs to write a few lines of code in order to implement a concrete program analysis task.



\subsubsection{Null-pointer Analysis}
Null pointer errors are a common type of error in programming practices, which occur when uninitialized pointers are used. These errors not only cause program crashes during runtime, severely affecting user experience, but may also lead to data loss or inconsistencies in program state\cite{zhang2023detecting}. 
Therefore, there is a need to automatically detect and thereby mitigate these errors before releasing the code to public.
However, it is non-trivial to automatically locate this kind of error, as it involves field-aware inter-procedural data flow analysis.

\begin{lstlisting}[caption={Sample code with an Null-pointer error.}, label={code:null pointer}]
class Property{ pp=1; }
class T{
    p: property;
    printP(){ console.log(this.p.pp); }
}
function Main(){
    let t1 = new T();
    t1.printP(); // null pointer error
}
\end{lstlisting}

For example, Listing~\ref{code:null pointer} illustrates an interprocedural null pointer error. 
In the \texttt{Main} function, \texttt{t1.p.pp} will be utilized. But in reality, \texttt{t1.p} is undefined at this point, which will cause the program to crash.



\begin{figure}[h]
    \centering
    \includegraphics[width=\linewidth]{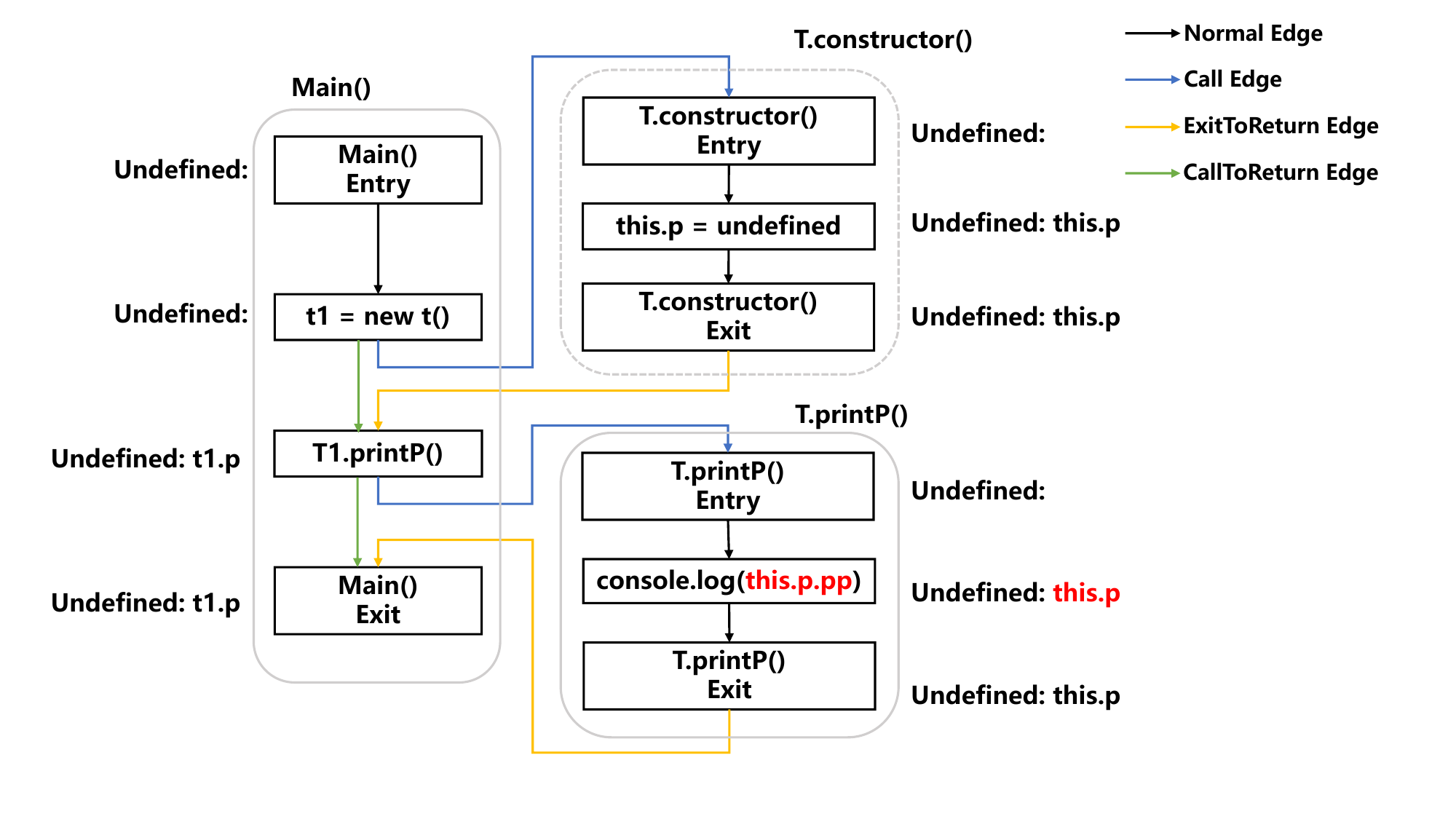}
    \caption{The process to implement Null-pointer detectors.}
    \label{fig:nullPointer}
\end{figure}

To facilitate the implementation of inter-procedural data-flow analyses, we have implemented in \toolname the famous IFDS (Interprocedural Finite Distributive Subset) algorithm\cite{reps1995precise, naeem2010practical}, which provides a flexible framework that allows developers to define data flow facts and transfer functions as needed.
Taking the aforementioned null-pointer error detection as an example, one only needs to extend the given IFDS framework to define how will the data propagate. 
Figure~\ref{fig:nullPointer} illustrates the handling process of the example code in Listing~\ref{code:null pointer}. 

Generally, the data propagation between statements is divided into four types of edges: Normal Edge, Call Edge, ReturnToExit Edge, and CallToReturn Edge. Each type of edge has a different data flow processing function. 
Ultimately, \toolname will accurately detect which line of code will cause a null pointer exception.


This example further demonstrates the usefulness of \toolname, being able to be leveraged to implement automated null pointer error detector. 

\section{Threats to Validity}

\paragraph{Internal threats to validity} The efficiency evaluation results (Section~\ref{sec:efficiency}) may be affected by other services running in the experimental environments (i.e., 64-bit Windows 11). Besides, in the accuracy evaluation(Section~\ref{sec:accuracy}), we employed a sampling approach to manually verify the invocation edges. This inherently carries the potential for inaccuracies.

\paragraph{External threats to validity} Considering that \OH is still in its early development stage, the features of \arkts may change a lot. Our proposed \toolname needs continuous updates in the future. Moreover, the dataset of \OH applications in our study was conducted up to April 2024. Given the rapid development pace of the HarmonyOS ecosystem, it is anticipated that the number of applications has significantly increased since then, and some applications in our dataset may have been updated. 

\section{Related Work}
\label{sec:relatedwork}

Static analysis has been regarded as one of the most important techniques in the field of software engineering\cite{li2017static, blackshear2018racerd}, assisting developers and researchers in security analysis\cite{arzt2014flowdroid, li2015iccta, rahaman2019cryptoguard}, vulnerability detection\cite{li2015potential, liu2021first, sun2021characterizing}, and so on.
To facilitate the development of static analysis approaches, our fellow researchers and practitioners have proposed to our community various static analysis frameworks.
Table~\ref{lab:frameworks} summarizes some of the representative frameworks grouped based on their targeted programming languages, such as Java~\cite{vallee2010soot,karakaya2024sootup,WALA,bravenboer2009strictly,tan2023tai}, C/C++\cite{llvm,lattner2008llvm,schubert2019phasar,sui2016svf}, JavaScript and Typescript\cite{jensen2009type,kashyap2014jsai,eslint}, Python\cite{li2022scalpel}, Swift\cite{tiganov2020swan}, and Rust\cite{li2024context}.

\paragraph{Program analysis tools} A large number of static analysis tools have emerged based on static analysis frameworks, such as FlowDroid for detecting sensitive data-flows\cite{arzt2014flowdroid}, CiD for detecting API-induced compatibility issues\cite{li2018cid}, IccTA for inter-component data flow analysis\cite{li2015iccta}, etc. These tools each focus on specific areas of code analysis, helping developers improve code quality and security.

\begin{table}[ht]
  \caption{The list of representative static analysis frameworks.}
  \label{lab:frameworks}
  \begin{center}
   \resizebox{1\linewidth}{!}{
    \begin{tabular}{c|c|p{0.7\linewidth}} 
     \toprule
\textbf{Language}  & \textbf{Framework} & \textbf{Paper Title Or GitHub Page}  \\\hline
\multirow{4}{*}{Java}  &  Soot/SootUp & Soot: A Java bytecode optimization framework \\  
~ &   WALA & https://github.com/wala/WALA \\   
~ &   Doop & Strictly declarative specification of sophisticated points-to analyses \\   
~ &  Tai-e & Tai-e: A developer-friendly static analysis framework for Java by harnessing the good designs of classics \\ \hline 
\multirow{2}{*}{C/C++}   &  SVF & SVF: interprocedural static value-flow analysis in LLVM \\ 
~ &  PhASAR & Phasar: An inter-procedural static analysis framework for c/c++ \\ \hline 
\multirow{3}{*}{JS/TS} &   TAJS & Type Analysis for JavaScript \\  
~ & JSAI & JSAI: a static analysis platform for JavaScript \\  
~ &  ESLint & https://github.com/eslint/eslint \\ \hline 
Python   & Scalpel & Scalpel: The python static analysis framework \\ \hline
Swift   &   Swan & Swan: A static analysis framework for swift \\ \hline
Rust & RUPTA & A Context-Sensitive Pointer Analysis Framework for Rust and Its Application to Call Graph Construction \\
 \bottomrule    
    \end{tabular}
    }
    \end{center}
\end{table}

\section{Conclusion}
\label{sec:conclusion}

In this work, we present the first static analysis framework \toolname for \OH Apps to the community.
\toolname addresses the problems of existing program analysis methods and has a set of common features (e.g., call graph construction)
that are recurrently required when implementing in-depth static analyzers such as privacy leak detectors and compatibility issue detectors.
We have collected and open-sourced a HarmonyOS native application dataset and conducted a series of evaluations on \toolname, confirming its high performance and accuracy, intermediate representation (IR) readability, and ease of use.
As for our future work, we commit to keep improving the \toolname framework so as to support our fellow researchers in implementing efficient tools to resolve realistic App analysis problems.






\bibliographystyle{unsrt}
\bibliography{main}

\end{document}